\documentclass[fleqn,10pt]{wlscirep}
\usepackage[utf8]{inputenc}
\usepackage[T1]{fontenc}
\usepackage{float}
\usepackage{xcolor}

\title{Non-contact photoacoustic imaging with a silicon photonics-based Laser Doppler Vibrometer}

\author[1,2,*]{Emiel Dieussaert}
\author[1,2]{Roel Baets}
\author[3]{Hilde Jans}
\author[3]{Xavier Rottenberg}
\author[1,2]{Yanlu Li}
\affil[1]{Photonics Research Group, Ghent University-IMEC, Technologiepark-Zwijnaarde 126, 9052 Ghent, Belgium}
\affil[2]{Center for Nano- and Biophotonics, Ghent University, Technologiepark-Zwijnaarde 126, 9052 Ghent, Belgium}
\affil[3]{IMEC, Kapeldreef 75, 3001 Leuven, Belgium}

\affil[*]{Emiel.Dieussaert@ugent.be}

\begin{abstract}
Photoacoustic imaging has emerged as a powerful, non-invasive modality for various biomedical applications. Conventional photoacoustic systems require contact-based ultrasound detection and expensive and bulky high-power lasers for the excitation. The use of contact-based detectors involves the risk of contamination, which is undesirable for most biomedical applications. While other non-contact detection methods can be bulky, in this paper, we demonstrate compact and contactless detection of photoacoustic signals on silicone samples embedded with ink-filled channels, mimicking tissue with blood-carrying veins. A silicon photonics-based Laser Doppler Vibrometer (LDV) detects the acoustic waves excited by a compact pulsed laser diode. By scanning the LDV beam over the surface of the sample, 2D photoacoustic images were reconstructed of the sample. Photoacoustic signals with absorption coefficients within the physiological range were detected by this setup.
\end{abstract}
\begin{document}

\flushbottom
\maketitle
\thispagestyle{empty}

\section*{Introduction}

Over the past decade, photo-acoustic imaging (PAI) has gained significant attention in biomedical fields \cite{Wang2012, Yao2013, Zhang2006, Xu2006}, including cancer detection \cite{Kim2010, Xu2006}, vascular imaging \cite{Kim2010, Li2021, Sun2022},  and functional brain imaging \cite{Na2022} amongst others. In this technique, the absorption of light by tissue chromophores causes local heating and expansion, which results in the emission of an acoustic wave. Detection of these acoustic waves allows the reconstruction of the absorption profile in the sample.  The combination of light for excitation and sound for detection allows for going beyond existing optical techniques, like optical microscopy or optical coherence tomography (OCT) \cite{Aumann2019}, in terms of imaging depth.

In conventional PAI systems, ultrasound waves generated through the photoacoustic effect are detected using contact-based detectors, typically ultrasound transducers \cite{Xu2006}. Due to the large impedance mismatch between sample and air and the high absorption of ultrasound in air, these detectors require direct contact with the sample of interest, often necessitating the use of coupling media such as gel or water to ensure efficient acoustic coupling \cite{Xu2020}. 

While contact-based detectors such as piezoelectric and capacitive micromachined transducers \cite{Qiu2015, Khuri2011} have proven to be effective for many applications, direct contact between the detector and a biomedical sample or patient, often facilitated by coupling gels, presents a risk of contamination \cite{Sartoretti2017}. Moreover, most detectors are opaque and hereby limit efficient excitation light delivery. While most research on ultrasonic transducers focuses on decreasing the size or increasing the sensitivity \cite{Westerveld2021}, our approach aims to remotely detect the ultrasound signals. 

Over the past decades, optical techniques have been used as a non-contact alternative detection method \cite{Wissmeyer2018}. One such promising approach involves the use of laser Doppler vibrometry (LDV). LDV is a well-established non-contact optical technique for measuring surface vibrations, and it has recently shown promise for the remote detection of the photoacoustic waves generated in PAI \cite{Tian2016, Wang2020}. 
This non-contact approach eliminates mechanical coupling artifacts, reduces the risk of sample damage, and enables the study of delicate biological samples.  

Photoacoustic imaging often requires detection of the acoustic waves at various locations on the surface of the sample. Most LDV systems, however, are limited to only a few detection beams and therefore require scanning the surface of the sample, creating a complex and expensive system and compromising the imaging speed.  A solution for this could be to create multi-beam LDVs. However, conventional LDVs are fiber- or free-space-based systems and use discrete components for each beam, which makes scaling the number of detection points bulky and expensive. 

Over the past few decades, silicon photonics has steadily gained recognition as a reliable platform for integrated optics, enabling the miniaturization and large-scale integration of optical components. The development of on-chip LDV systems based on silicon photonics has the potential to overcome the drawbacks of current LDV systems, providing compact and relatively cheap multibeam LDVs \cite{Li2018, Li2013}.

The primary objective of this work is to demonstrate the feasibility of using a silicon photonics-based LDV system as a detector for photoacoustic measurements and to assess the limitations of this approach. Additionally, we aim to employ a compact laser diode (LD) source as the photoacoustic excitation source, which will contribute to reducing the overall size and cost of the setup \cite{Singh2020}. 

\section*{Methods}
\begin{figure}[ht]
\centering
\includegraphics[width=14 cm]{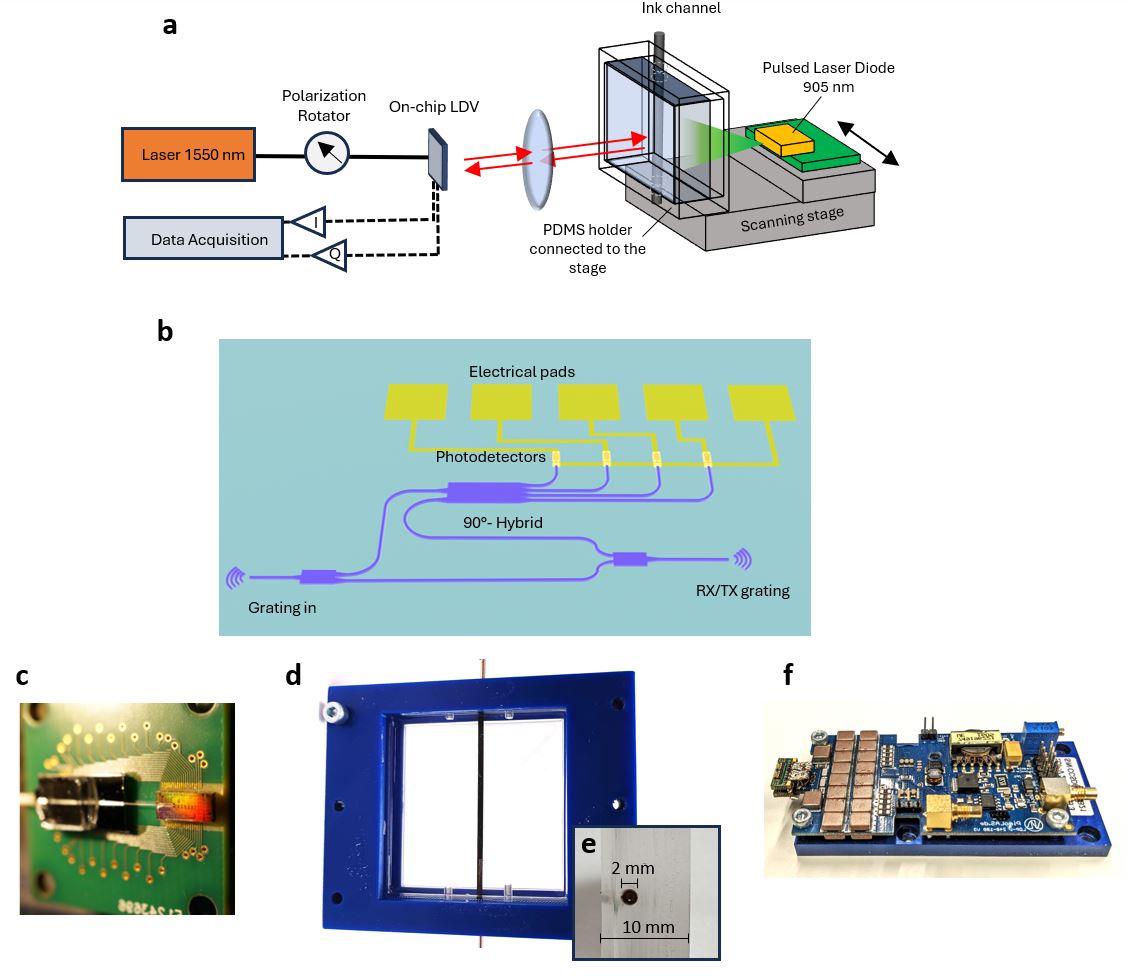}
\caption{a) A photoacoustic setup consisting of an on-chip LDV with a 1550~nm laser, connected to a data acquisition module. The probe beam is delivered by an optical lens system to the surface of the sample. On the right side of the sample, the pump laser diode fires optical pulses toward the silicone sample where the ink in the channel absorbs the light, hereby generating acoustic waves. b) Schematic of an on-chip homodyne LDV where light is entering the chip through 'Grating in' and after splitting into a reference and measurement arm is probing the target using the RX/TX grating. Reference and probe light are combined in a 90-degree optical hybrid. c) Picture of the optical chip wire-bonded to an interposer PCB and an optical fiber glued to the input grating. d) and e) Pictures of the transparent PDMS with an ink-filled embedded channel and its cross-section. f) A picture of the 905 nm laser diode connected to the Picolas pulsed laser driver.}
\label{fig:system}
\end{figure}

A photoacoustic system can be divided into three main elements; first of all the excitation light source, secondly the sample converting the excitation light into an acoustic signal and lastly an acoustic detection method. Figure \ref{fig:system} a, shows a schematic of the setup used in this paper.
On the left side of the sample in \ref{fig:system}a., there is the contactless detection system with the probe beam directed toward the sample, consisting of a laser source, the photonic integrated circuit (PIC), and the data acquisition module.
On the right side of the sample, a 905~nm pulsed laser diode acts as the photoacoustic excitation source directed towards the silicone sample with an embedded ink channel that absorbs the excitation light and mimics the acoustic properties of a biological sample.

The photoacoustic signals generated from the ink channel travel through the silicone toward the sample's surface, inducing small vibrations on the surface, which are detected by the chip-based  LDV. To detect the small photoacoustic signals, averaging over multiple excitations was enabled using the triggered acquisition of LDV-detected signals synchronized with the firing of the pulsed diode. 

As mentioned before, photoacoustic imaging requires the detection of the signals at multiple locations. Although photonic integration could allow for the dense integration of multi-beam LDVs, for this paper a single-beam LDV was used while moving the sample and laser diode, both attached to the same scanning stage.
\subsection*{Silicon Photonics-based Laser Doppler Vibrometer}

A homodyne LDV is an optical interferometric method to measure vibrations; the combination of reference light and measurement beam on a photodetector results in a current depending on the relative phase of both beams.  In this paper, we use a Silicon-On-Insulator platform to develop an integrated homodyne  LDV. The high index contrast allows for the miniaturization of many optical components, fabricated using CMOS-compatible techniques. Here, we use IMECs ISIPP50G platform to integrate most of the optical components after the coupling of light \cite{Siew2021}. A fiber-coupled DFB~laser diode delivers 1550~nm-light to the PIC through a fiber glued to couple into the input grating coupler using a polarization rotator to ensure correct polarization into the chip.  Figure \ref{fig:system}b, shows a schematic of the on-chip LDV which acts as a Mach-Zehnder interferometer.  On the PIC, the optical power is split into a measurement and reference path. The light in the measurement waveguide is directed towards a transmit/receive grating coupler (RX/TX grating). There, it is coupled out from the chip and, using a lens system, focused on the target.  After reflection from the target, it is coupled back into the chip.  Using an optical power meter at the position of the target, it was measured that around 0.5~mW is coupled from the chip to the target. This is within the eye-safety limits at this wavelength according to the ANSI standards. This probe light is then combined with the reference light in a 90-degree- optical hybrid on the PIC \cite{Halir2011}.  The 4 outputs of the hybrid contain a combination of the measurement and detection light, each time with a different 90-degree relative phase shift. On-chip photodetectors convert the optical signals into two current pairs. The electrical pads on the PIC are wire-bonded to a printed circuit board (Fig. \ref{fig:system}), where differential amplification of these current signal pairs, results in an in-phase (I) and quadrature (Q) - electrical signal. After amplification, the I and Q signals are recorded using a data acquisition card (Gage; both channels have a sampling rate of $65 \times 10^6$ samples per second). From these recorded signals, the relative phase difference $\Delta \theta(t)$ between the reference and measurement beam can be demodulated using Eq. \ref{eq:demod}, and as such, relative movements of the target can be detected. 
\begin{equation}
\label{eq:demod}
    \Delta\theta(t) = \arctan(\frac{Q(t)}{I(t)})
\end{equation}
Note that for a perfect system, the I and Q points constitute a perfect circle when inducing large vibration amplitudes (> 1550~nm), however in reality imperfections in fabrication result in DC offsets and an elliptical shape, which can be compensated for as discussed in the section 'Photoacoustic signal processing and imaging' or using other approaches \cite{Li2023}.

From the phase difference  $\Delta \theta(t)$ obtained after demodulation, the target displacement can easily be calculated, with $\lambda= 1550~nm$, the wavelength of the probe beam.
\begin{equation}
\label{eq:disp}
    \Delta d(t) = \frac{\lambda \Delta \theta(t)}{4\pi}
\end{equation}

\subsection*{Pulsed Laser diode}
Conventional photoacoustic systems often use bulky and expensive pulsed laser sources to generate high-power nanosecond pulses. However, to enable compact photoacoustic imaging, we use a pulsed laser diode as a small and cheaper alternative. A commercially available pulsed laser driver (Picolas LDP-V 240-100) drives the pulsed laser diode (Osram, SPL S4L90A) with a wavelength of around 905~nm, generating optical pulses having a duration between 100 and 500~ns with a peak optical power of 500~ Watt, resulting in pulse energies ranging between 50 and 250~$\mu J$. Although the pulse energy of the pulsed laser diode is around one or two orders of magnitudes lower compared to more expensive pulsed laser systems like laser diode pumped optical parametric oscillators (OPO), a pulsed laser diode allows for higher repetition rates, which allows averaging the signal more, hereby partially compensating the lower signal \cite{Singh2020}. In this paper, the pulsed laser is operated at a repetition rate of 1~kHz and a pulse width of around 400~ns resulting in around 0.2~W average power and around 200~$\mu J$ of pulse energy, which is within the ANSI skin exposure safety limits.

\subsection*{Silicone sample}
Photoacoustic signals are generated through the sudden absorption of the laser light in a sample, causing local heating and expansion leading to the emission of a propagating pressure wave. In this paper, a silicone sample was developed featuring a channel filled with an absorbing solution. By putting an ink-water solution in this channel, the 905~nm light is absorbed there and generates photoacoustic signals upon pulsed laser illumination. The silicone samples were made by pouring Polydimethylsiloxane (PDMS, Sylgard 184) in molds with a thickness = 10-13~mm, containing rods with a diameter of 2~mm. After 12 hours of curing at 40 degrees Celcius, these rods were removed, resulting in an empty channel. The PDMS has a typical speed of sound of  1020 m/s \cite{Xu2020,Genoves2023}

It is important to note that the base of the mold was a glass surface to enable a smooth and specular reflective surface to enable efficient LDV detection. Various concentrations of water-based ink solutions (0.01~\% - 1~\%~Black India ink) were employed into the channel to act as the absorber and thus the origin of the photoacoustic signals. The absorption of the 0.1~\% ink solution was measured to have an extinction coefficient of 12.5 $cm^{-1}$.

\subsection*{Photoacoustic signal processing and imaging} \label{ssec:pa_processing}
Due to imperfections in the photonic circuit and electrical amplification, (resulting from an imbalance in the hybrid/photodetector or electrical amplifier circuit), the detected IQ points did not constitute a circle but rather an ellipse with a DC offset. To enable accurate demodulation, a reference measurement of a large vibration was recorded with the LDV before each measurement. This enables the fitting of the ellipse which then can be used to project all data points onto the unit circle before arctan demodulation as seen in Eq. \ref{eq:demod}, to enable accurate demodulation.  An example of a recorded and fitted ellipse can be seen in figure \ref{fig:data}a.

The sample and pulsed laser diode were attached to a scanning stage. Scanning the LDV beam on a line over the surface of the sample allows a 2D photoacoustic reconstruction of the deposited energy by the excitation laser.  A scanning pitch of 125~$\mu m$  was used and the demodulated signals were averaged over 1~second for each location and low-pass filtered with a cutoff frequency of 5~MHz to limit the noise. As stated in Eq. \ref{eq:disp}, the displacement can be calculated using the demodulated signal. Then, by taking the temporal derivative an estimate of the velocity $u$ of the sample's surface can be retrieved. The pressure $p$ measured at the surface can be estimated using the following relation with the acoustic impedance of the sample $Z_s>>Z_{air}$ \cite{Rousseau2012}:

\begin{equation}
    p = \frac{Z_s}{2} u
    \label{eq:pressure}
\end{equation}

To reconstruct the image we used a 2D time-reversal algorithm using MATLAB package k-wave \cite{Treeby2010}. The algorithm uses the measured forward-propagation field to generate back-propagation fields to reconstruct initial pressures at time = 0. For the image reconstruction in this paper, we used a reported speed of sound in PDMS samples of 1020 m/s  \cite{Xu2020,Genoves2023}. 

\section*{Results and Discussion}
\begin{figure}[h!]
\centering
\includegraphics[width=14cm]{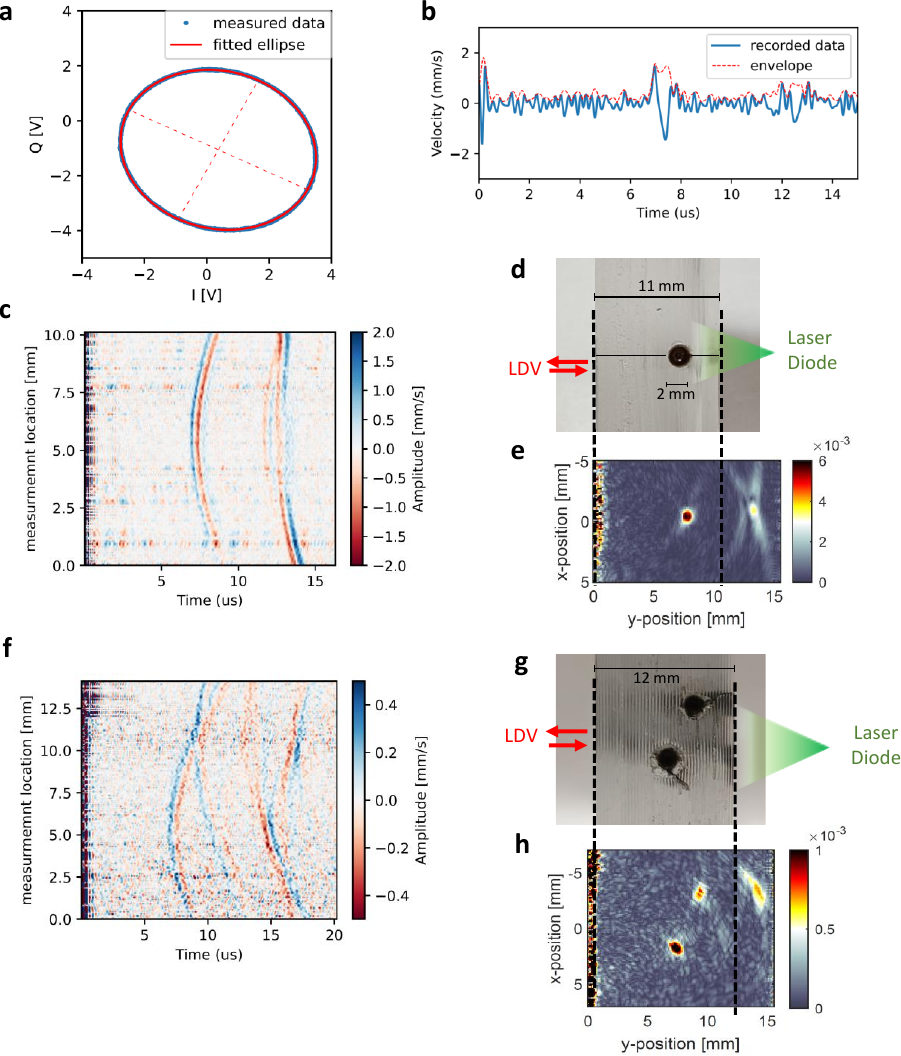}
\caption{a) Reference measurement of the I and Q signals and fitting of the ellipse before demodulation. b) Time trace of a recorded and demodulated photoacoustic signal. c) Plot of the photoacoustic signal for different locations along the scanning direction of the probe beam on the surface of the sample depicted in d). e) Time reversal reconstruction of d) using the data of c). f) scanning data for a dual channel sample (depicted in g). h) Time reversal reconstruction for the dual channel sample.}
\label{fig:data}
\end{figure}

Fig.\ref{fig:data}~b shows a time trace of a demodulated and averaged photoacoustic signal recorded by the on-chip LDV on the single channel sample (as depicted in fig. \ref{fig:system}  d and \ref{fig:system}e ), with the channel containing a 1~\% ink-water solution. To record this image, all segments over 1~second were averaged (1000 times) after demodulation.  Time equals zero indicates the firing of the pulse and shows an interference with the detection system. After a period of around 7~$\mu s$, a clear movement of the surface is detected by the LDV. The delay between excitation and signal arrival tells the distance between the signal origin and the detection location. 

In order to create a 2D photoacoustic image, the LDV-beam was scanned along a line on the surface of the sample. Figure~\ref{fig:data}~c shows the detected photoacoustic signals at different locations for an embedded light-absorbing channel (depicted in figure \ref{fig:data}~d). With a scanning pitch of  $125~\mu m$, well below the cutoff acoustical wavelength ($\lambda_c \approx 200~\mu m$), and a total distance of around ~$1~cm$, signals were recorded at 80 different locations. From figure \ref{fig:data}~c, it can be seen that for some locations the signal-to-noise ratio is better compared to others. This is due to the varying collection of optical power after reflection from the sample surface due to impurities or dust on the detection surface. It has to be noted that for biomedical applications where the surface sample is rough and has diffuse reflectivity,  high-NA lenses would be necessary for the LDV probe beam to ensure adequate collection of light. Figure~\ref{fig:data}~c  shows a clear relation between the measurement location and signal delay, but also a secondary signal originating from the reflection of the generated acoustic signal by the backside of the sample. 

Using the time reversal algorithm, mentioned in the methods section, a reconstruction image can be made (Fig.~\ref{fig:data}e). This image shows the correct location of the embedded channel relative to the detection locations and shows a mirror image due to the reflection from the backside of the sample. Fig.~\ref{fig:data}f and e show the detected signals for a sample with two channels embedded (\ref{fig:data}g). To illuminate both channels, the laser diode was placed further away from the sample, resulting in weaker illumination and thus lower signal generation.  

\begin{figure}[h!]
\centering
\includegraphics[width=11cm]{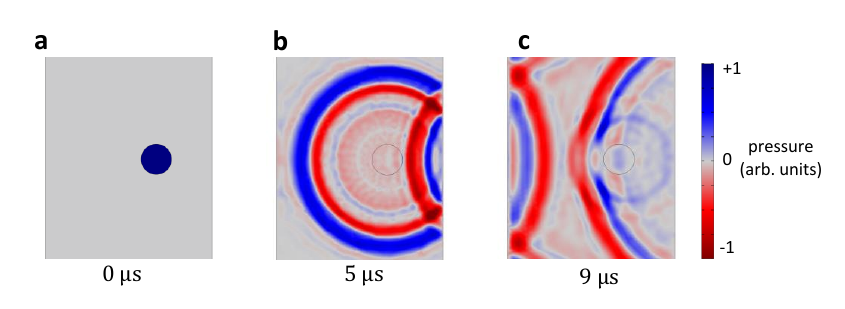}
\caption{ 2D COMSOL simulation of acoustic propagation in a geometry similar to Fig. \ref{fig:system} e for an initial pressure distribution with increased pressure inside the tube filled with a water-based solution surrounded by PDMS at a) $t=0~\mu s$, b) $t=5~\mu s$ and c) $t=9~\mu s$}
\label{fig:sim}
\end{figure}

From Fig. \ref{fig:data}c, it can also be noted that the reflected signal has a different shape compared to the primary signal. This secondary signal contains some 'splitting' of the acoustic pulse into two contributions arriving right after each other for the measurements locations near the norm of the channel. A simulation using COMSOL of the scalar wave-equation of a simplified acoustic propagation problem reveals that the splitting originates from the transmission of the reflected signal through the channel (Fig. \ref{fig:sim}). The simulation assumes a uniform acoustic impedance of the PDMS of around  1.02 MRayl \cite{Xu2020,Genoves2023}, an acoustic impedance of the channel of 1.5 MRayl and a uniform increased initial pressure across the channel (Fig. \ref{fig:sim} a).  The geometry is similar to the dimensions from figure \ref{fig:data} d and soft sound boundary conditions are assumed for the PDMS/air interface.  After propagation for around $5~\mu s$, Fig. \ref{fig:sim} b shows that the primary signal is propagating and almost reaching the detection surface (left), while the secondary channel has appeared due to the reflection from the backside. Note that after $5~\mu s$, no splitting of the reflected signal can be observed. However, the shape of the reflected signal after it has passed by the channel, as can be seen in Fig. \ref{fig:sim} c, shows the splitting of the reflected signal. The first contribution of this reflected wave is the signal passing through the channel with a higher speed of sound while the second contribution signal is the diffracted signal going 'around' the channel.

\begin{figure}[h!]
\centering
\includegraphics[width=12 cm]{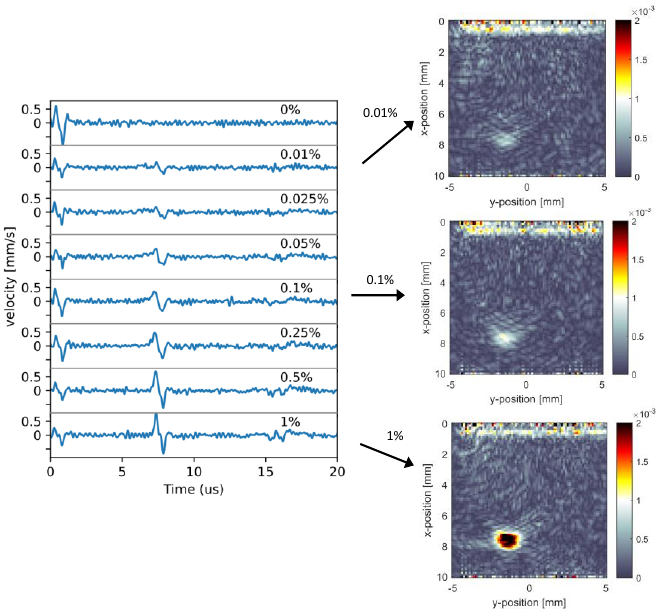}
\caption{On the left, signal time traces of the recorded velocity of the surface after photoacoustic excitation for different ink concentrations inside the sample. Increasing the concentrations results in stronger signals. The 0.1 \% ink solution was measured to have an absorption of 12.5 $cm^{-1}$. On the right, we show image reconstructions for different ink concentrations showing reduced contrast for reduced concentrations.}
\label{fig:concentrations}
\end{figure}

Figure \ref{fig:concentrations}a, shows signal amplitudes depending on the absorption of the ink solution for 10~second averaging. Higher ink concentrations show high signal generation and clear photoacoustic images. As can be seen, we detect small photoacoustic signals down to 0.01~\% which agrees with an absorption coefficient of around 1.2 $cm^{-1}$, which is below the typical physiological range of blood absorption for a wavelength of 905~nm ($2- 10~cm^{-1}$)   \cite{Yao2013}. It has to be noted that due to the large range of measured concentrations, the measured peak velocity does not scale linearly, which could be the result from saturation effects and having a bandwidth that is too small to capture very short pulses.

As mentioned before, the sensitivity of the LDV is largely dependent on the captured reflected power.  To record the data in figure \ref{fig:concentrations}, the captured reflected optical power was optimized at a single location to have a good reflection.  Using the data shown in figure \ref{fig:concentrations}, the noise on the velocity can be calculated by taking the root mean square error (RMSE) off the recorded data for 0~\% ink after 2 $\mu s$ (to remove the spurious influence of the laser excitation), resulting in a RMSE of around 0.08~mm/s. Using Eq.\ref{eq:pressure} with an impedance of 1.02 MRayl, the RMSE is $p_{rmse} = 80~Pa$. To calculate the noise equivalent pressure (NEP) \cite{Yao2014}, we need to take into account a factor of 100 to compensate for the 10~second averaging, resulting in an $NEP = 0.8 kPa$. Although these results show comparable performance compared to commercial interferometric-based techniques \cite{Tian2016}, contact-based photoacoustic systems with state-of-the-art detectors have better noise equivalent pressures for larger bandwidths \cite{Wissmeyer2018}. Considering the bandwidth of 5 MHz of the LDV-based detection system and assuming an impedance of 1.02~MRayl, the estimated resolution of the photoacoustic images is limited to $0.8\lambda_{c}\approx160~\mu m$, with $\lambda_{c}$ being the acoustic wavelength for the cutoff frequency.\cite{Beard2011, Xu2003}. 

\section*{Conclusion and Outlook}

In this paper, a photonics-enabled homodyne LDV was developed to act as a compact and contactless method for photoacoustic detection.  We demonstrated contactless photoacoustic imaging by scanning the probe beam over the sample's surface while using a small and cheap pulsed laser diode as the excitation source. Although the demonstration in this paper only uses a single on-chip LDV beam, therefore requiring scanning along the surface to enable photoacoustic imaging, photonic integration allows for scaling multi-beam LDV layouts, which could omit the need for scanning. The samples used in this paper were clear, flat PDMS-based samples with an ink solution-filled channel embedded, acting as the absorber. Photoacoustic signals could be detected for ink solutions with an absorption within the physiological range for blood \cite{Yao2013}. Considering biomedical applications where blood acts as a contrast agent, future implementations could include multi-wavelength excitation which could yield quantitative information on the local oxygenation \cite{Zhang2006}.  In summary, this work demonstrates the potential of silicon photonics-enabled LDVs for compact and contactless ultrasound detection for photoacoustic imaging.

\section*{Data Availability Statement}
The dataset and signal processing algorithms are available on \href{https://github.com/edieussa/PA_Data_and_Figures}{github.com/edieussa/PA\_Data\_and\_Figures} or will be available from the corresponding author upon request.

\section*{Acknowledgments}
The authors thank Sheila Dunphy for helping with the development of the PDMS sample, Hendrik Spildooren, Mohammadamin Ghomashi, and Hasan Salmanian for helping with the electronics around the PIC and excitation laser, Selènè Spruytte for the development of the first versions of the sample and Steven Verstuyft and Bjorn Vandecasteele for wire-bonding the PIC. Yanlu Li acknowledges European Commission Horizon 2020 grant number 101016738 (PhotonicLEAP)

\section*{Author contributions statement}

E.D,  R.B., H.J., X.R., Y.L. conceived the experiment.
E.D. and Y.L. built the on-chip system and photoacoustic setup and conducted sensing and imaging experiments. 
E.D,  R.B., H.J., X.R. and Y.L. analysed the results.  All authors reviewed the manuscript. 

\bibliography{export}

\end{document}